\documentclass[useAMS]{mn2e}
\usepackage{graphicx}
\usepackage{multirow}
\usepackage{txfonts}
\usepackage{color}

\title[Pulsational instability in B-type supergiant stars]{Pulsational instability in B-type supergiant stars}
\author[J. Daszy\'nska-Daszkiewicz, J. Ostrowski, A. A. Pamyatnykh]
{J. Daszy\'nska-Daszkiewicz$^{1}$\thanks{E-mail: daszynska@astro.uni.wroc.pl}, J. Ostrowski$^{1}$\thanks{E-mail: ostrowski@astro.uni.wroc.pl},
A. A. Pamyatnykh$^{2}$\thanks{E-mail: alosza@camk.edu.pl}\\
$^{1}$Instytut Astronomiczny, Uniwersytet Wroc{\l}awski, ul. Kopernika 11, 51-622 Wroc{\l}aw, Poland\\
$^{2}$Nicolaus Copernicus Astronomical Center, Bartycka 18, 00-716, Warsaw, Poland\\
}
\begin{document}

\date{Accepted ... Received ...; in original form ...}

\pagerange{\pageref{firstpage}--\pageref{lastpage}} \pubyear{2002}

\maketitle

\label{firstpage}

\begin{abstract}
We present results of the instability analysis of the post-main sequence massive star models against radial and nonradial pulsations.
We confirm that both p- and g-modes can be excited by the $\kappa$-mechanism acting in the metal opacity bump.
However, as opposed to the previous claims, we find that an intermediate convective zone (ICZ) related to the hydrogen burning
shell is not necessary for excitation of g-modes. These modes can be reflected at a minimum of the Brunt-V\"ais\"al\"a frequency,
located at the top of the chemical composition gradient region surrounding the radiative helium core.
This minimum is associated with the change of actual temperature gradient
from the adiabatic value in the semiconvective zone to the radiative value above it.
Thus, the existence of pulsations at this evolutionary stage does not prove
the existence of the convective zone but only some reflective layer.
Finally, we show that no regular patterns can be expected in oscillation spectra of blue supergiant pulsators
but there is a prospect for identification of the mode degree, $\ell$, from multicolour photometry.
\end{abstract}

\begin{keywords}
stars: early-type -- stars: supergiants -- stars: oscillations
\end{keywords}

\section{Introduction}
The properties of pulsations of the main sequence B-type pulsators such as $\beta$ Cephei stars and Slowly Pulsating B-type
stars (SPB) have been studied extensively for years, but very little is known about pulsational properties of stars after Terminal
Age Main Sequence (TAMS). Moreover, until recently no pulsating post-SPB star has been found.
Using the data from the MOST satellite for the blue supergiant HD 163899 (B2 Ib/II, Klare \& Neckel 1977, Schmidt \& Carruthers
1996), Saio et al. (2006) found 48 frequencies with the values less than 3 c/d and maximum amplitudes of a few millimagnitudes.
The authors attributed these light variations to the g- and p-mode pulsations.
HD 163899 is a less luminous star than $\alpha$ Cyg variables (van Leeuwen et al. 1998) and the first of previously unknown type of variable,
i.e., Slowly Pulsating B-type supergiants, (SPBsg), as termed by Saio et al. (2006).

According to previous calculations (Pamyatnykh 1999; Pamyatnykh \& Ziomek 2007) the instability domain of SPB stars was perfectly confined
to the main sequence and the instability band of $\beta$ Cephei stars was slightly
extended behind TAMS. Saio et al. (2006) examined pulsational instability of models in the mass range of
$7 \le M/M_\odot \le 20$. The most important result of their analysis was to find the large g-mode instability domain in
post-main sequence models. In the hotter part of this instability region, p-modes are also excited. It is in agreement with
stellar parameters estimated for HD 163899 which locate the star in this instability band.
Moreover, according to computations of Saio et al. (2006), the frequency range of HD 163899 covers the g- and p-mode pulsations.

Before the discovery of pulsational modes in HD 163899 it had been believed that there is no possibility for g-modes to be
excited in B-type supergiants. It is due to a strong radiative damping which occurs in a dense radiative helium core
due to the very large value of the Brunt-V\"ais\"al\"a frequency. Saio et al. (2006) have explained excitation of g-modes
in such stars owing to a partial reflection of some pulsational modes at an intermediate convective zone (ICZ) related to the hydrogen
burning shell. Thus, ICZ can prevent some modes from entering the radiative core and $\kappa$-mechanism in the superficial
layers might be effective. This result has been confirmed by Godart et al. (2009) who studied also
the effects of mass loss and overshooting from the convective core during the main sequence phases.
Lebreton et al. (2009) examined the effect of the adopted convective instability criterion.
All these authors claimed that formation and properties of ICZ was crucial for a presence of unstable modes in SPBsg stars.
However, in this paper, we show that pulsational modes in the B-type supergiant models can be reflected at the chemical
composition gradient formed above the radiative helium core and a presence of ICZ in the hydrogen burning shell is not
{\it condicio sine qua non}.

Here, we are not going to interpret the oscillation spectrum of HD 1643899 because the effective temperature, luminosity and
rotational velocity of the star are very poorly determined or unknown and we postpone this task for the future when better
observations will be collected.

The structure of the paper is as follows. In Section\,2, we present the propagation diagram for a representative model and domains
of pulsational instability for SPBsg stars on the HR diagram. Properties of instability parameter and kinetic energy of modes
are described in Section\,3. Section 4 is devoted to the photometric diagnostic diagrams as tools for identification of the SPBsg modes.
The last section contains Conclusions.

\section{Excitation of oscillation modes in B-type supergiants}

Regions of a star where pulsation modes can propagate in radial directions are strictly defined by the values of the Lamb ($L_{\ell}^2$)
and Brunt-V\"ais\"al\"a ($N^2$) frequencies, which are give by
$$L^2_\ell=\frac{\ell(\ell+1) c_s^2}{r^2}, \eqno(1)$$
and
$$N^2\simeq \frac{g^2 \rho}{p} \left[ \frac{4-3\beta}{\beta} \left(\nabla_{\rm ad}-\nabla\right) +\nabla_\mu \right], \eqno(2)$$
respectively. In the above formulae, $\ell$ is the mode degree, $c_s$ is the sound speed, $\nabla$ is the temperature gradient and $\nabla_{ad}$ its adiabatic
value, $\nabla_\mu$  is the mean molecular weight gradient, and $\beta$ is a ratio of the gas pressure to the total pressure.
Other symbols have their usual meanings.
Wave propagates radially with the frequency $\omega$ only if $\omega^2 > L^2_\ell, N^2$ (p-modes) or $\omega^2 < L^2_\ell, N^2$ (g-modes)
and they are evanescent in the other cases. This results from the dispersion relation which relates the pulsational freqeuncy
with the radial wavenumber.
As can be easily seen from the above formulae, the Lamb frequency depends mostly on the mode degree and the distance from
the center, whereas the Brunt-V\"ais\"al\"a frequencies on the stellar structure details; in particular, on a presence
of the chemical composition gradient and convection.

Here, we will consider B-type supergiant models in the hydrogen burning phases. Such star is made up of the radiative helium
core, the hydrogen burning shell and extended envelope. The other possibility is that these objects have already entered the
core helium burning phase but it is not essential for the point.

The evolutionary models were computed using the Warsaw-New Jersey code.
We adopted both the OP (Seaton 2005) and OPAL (Rogers \& Iglesias 1996) opacity tables.
The AGSS09 chemical mixture (Asplund et al. 2009) was used with the hydrogen and metal abundance by mass of $X=0.7$ and $Z=0.02$, respectively.
In the Warsaw-New Jersey code the Ledoux criterion for a convective instability is applied, which states that convection appears if $N^2<0$.
In unstable layers the adiabatic temperature gradient is assumed but the element abundance is frozen. Justification for this procedure was given by Dziembowski (1977).
All our computations were done in the framework of the zero-rotation approximation, i.e., all effects of rotation on stellar structure
and pulsations were neglected. Moreover, no overshooting  from the convective core during the main sequence phase was allowed.

An example of the propagation diagram for a model with a mass of $16 M_\odot$ is shown in Fig.\,1. The model was computed
with the OP opacities and has effective temperature of $\log T_\mathrm{eff}=4.340$ and luminosity of
$\log L/L_\odot=4.730$. Due to high mass concentration in the radiative core, the value of $N$ is very large in the center.
The Lamb frequency is plotted for the mode degree $\ell=$1, 2. Both frequencies are in units of [c/d].
The radiative helium core is very dense and of a small size; its border is at the temperature of about $\log T=7.6$ which corresponds
to about 4\% of the stellar radius. Just above the core, we have the chemical gradient zone, whereas convection connected with
the hydrogen shell burning has not been developed because $N^2>0$ in this region. This is due to the fact that in regions of
nuclear burning the values of $\mu$ increases with increasing depth and the gradient $\nabla_\mu$ contributes
positively to $N^2$. Dashed strips show the frequency ranges of unstable modes with $\ell=0,1,2$. It is worth to note that in this model
the radial fundamental mode is unstable whereas the higher overtones are stable.
\begin{figure}
\begin{center}
 \includegraphics[clip,width=85mm,height=68mm]{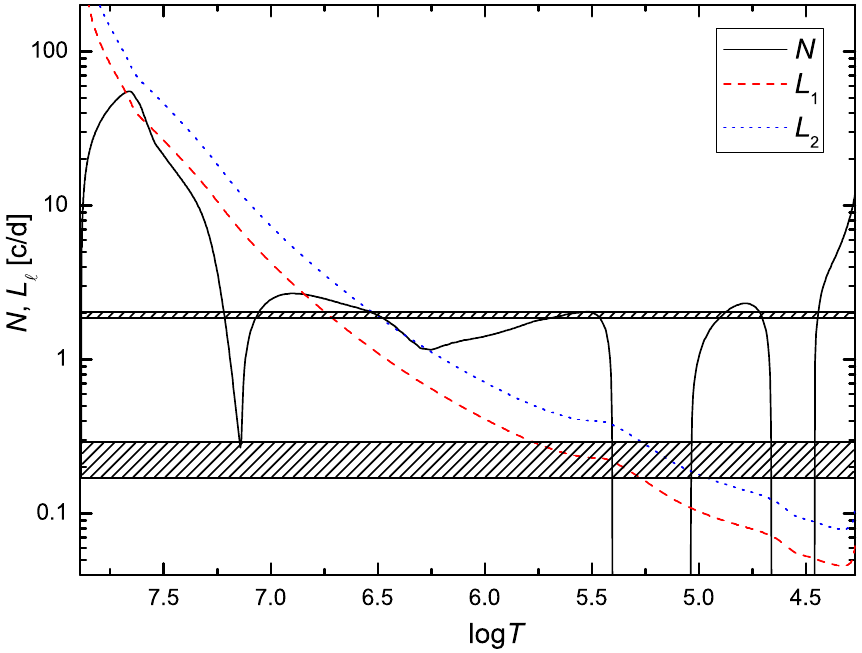}
   \caption{Runs of the Brunt-V\"{a}is\"{a}l\"{a} frequency, $N$ (solid line), and the Lamb frequencies, $L_\ell$, for $\ell=1,2$ (dashed and dotted lines respectively) in [c/d] for a star in the supergiant phase with a mass of $M=16 M_\odot$, $\log T_\mathrm{eff}=4.340$ and $\log L/L_\odot=4.730$. The model was computed with the OP opacities
   and AGSS09 solar mixture was adopted. The frequency ranges of unstable modes with $\ell=0,1,2$ are marked as dashed stripes.}
  \label{fig1}
\end{center}
\end{figure}
\begin{figure}
\begin{center}
 \includegraphics[clip, width=85mm]{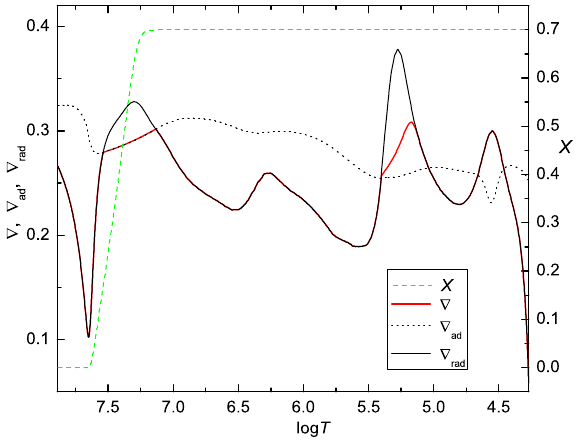}
   \caption{The hydrogen profile and runs of the temperature gradients in the model considered in Fig.\,1.}
  \label{fig22}
\end{center}
\end{figure}
A model with similar parameters but computed with the OPAL data has a similar overall run of $N$. The difference is that the OPAL model
reaches the lower minimum of $N$ around $\log T=7.1$ and has thinner convective layer associated with the Z-bump.
To give a better insight into the internal structure of the considered model, in Fig.\,2, we plotted the hydrogen profile, $X$,
and three temperature gradients: actual ($\nabla$), adiabatic ($\nabla_{\rm ad}$) and radiative ($\nabla_{\rm rad}$).
As we can see, almost in the entire interior, we have $\nabla=\nabla_{\rm rad}$, i.e., the whole energy is transported by radiation.
Only in the Z opacity bump at $\log T\approx 5.2$, convection is quite effective and carries up to 25 \% of the flux energy.
Our evolutionary models have been computed with the value of the mixing-length parameter, $\alpha_{\rm MLT}$, equal to 1.8.
The choice of this value influences the quantitative results on the pulsational instability, but qualitatively the results remain very similar,
as we tested using the models computed with $\alpha_{\rm MLT}=0.5$.
In the chemical gradient zone (semiconvective zone), we have $\nabla_{\rm rad} > \nabla_{\rm ad}$ but it does not mean that some energy must be
carried by convection. As it was discussed by Dziembowski (1977), the required part of energy can be carried by short-wave nonadiabatic oscillations.
Mixing of chemical elements is not associated with this process. Therefore in the stellar model calculations we use the simplest description
of the semiconvective zone structure, i.e., we assume $\nabla=\nabla_{\rm ad}$ and do not mix the matter in this zone. Convection here is stabilized
by the hydrogen abundance gradient. Therefore, the Brunt-V\"ais\"al\"a ($N^2$) frequency is greater than zero there.
Note also that a local minimum of the radiative gradient at $\log T$ of about 7.6 is connected with inner boundary of the hydrogen burning shell,
where luminosity quickly increases outwards. Such a feature does not influence dynamical properties of the interior which we discuss in the paper.
\begin{figure*}
\begin{center}
 \includegraphics[clip,width=65mm,height=85mm,angle=-90]{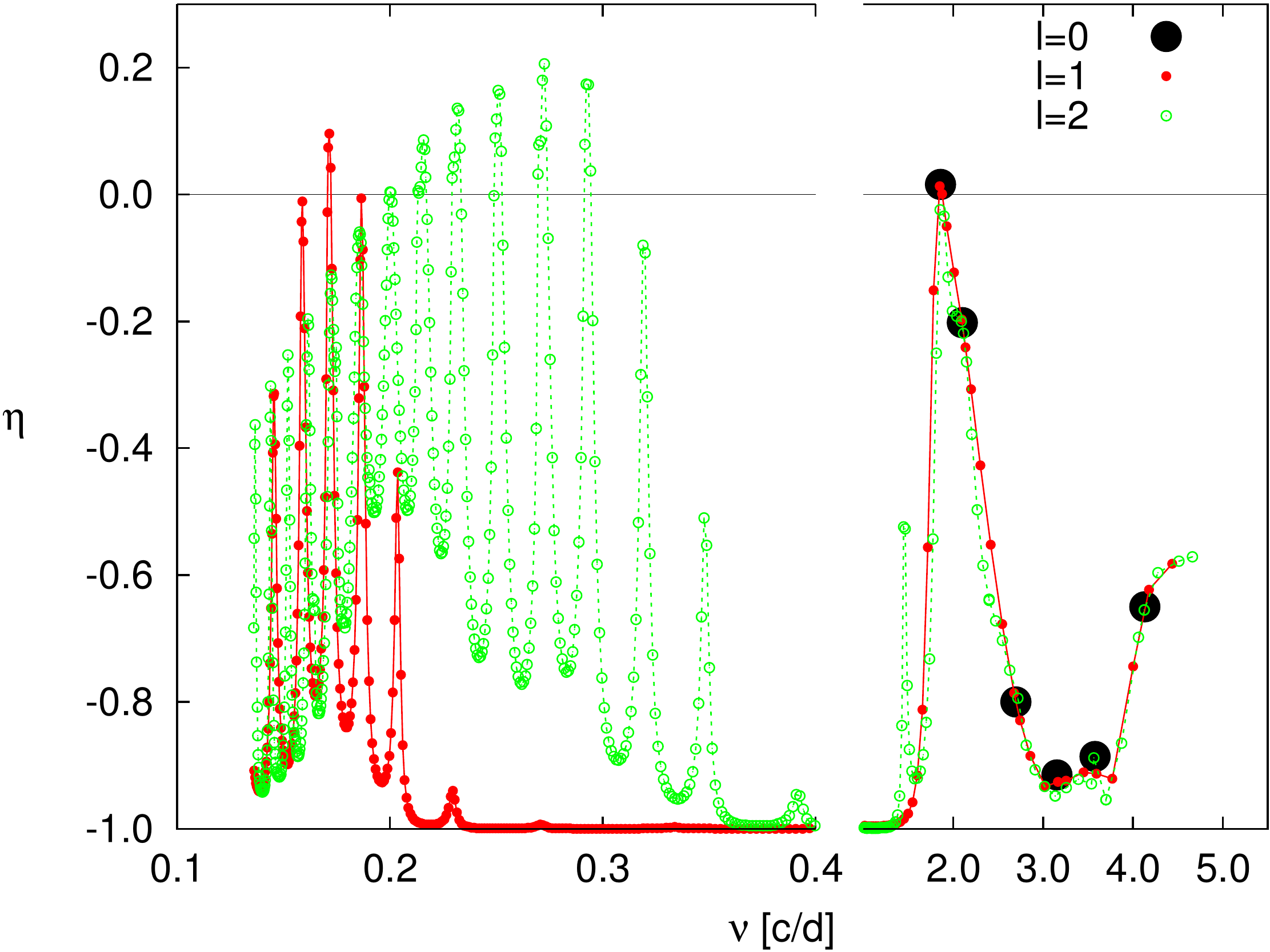}
 \includegraphics[clip,width=65mm,height=85mm,angle=-90]{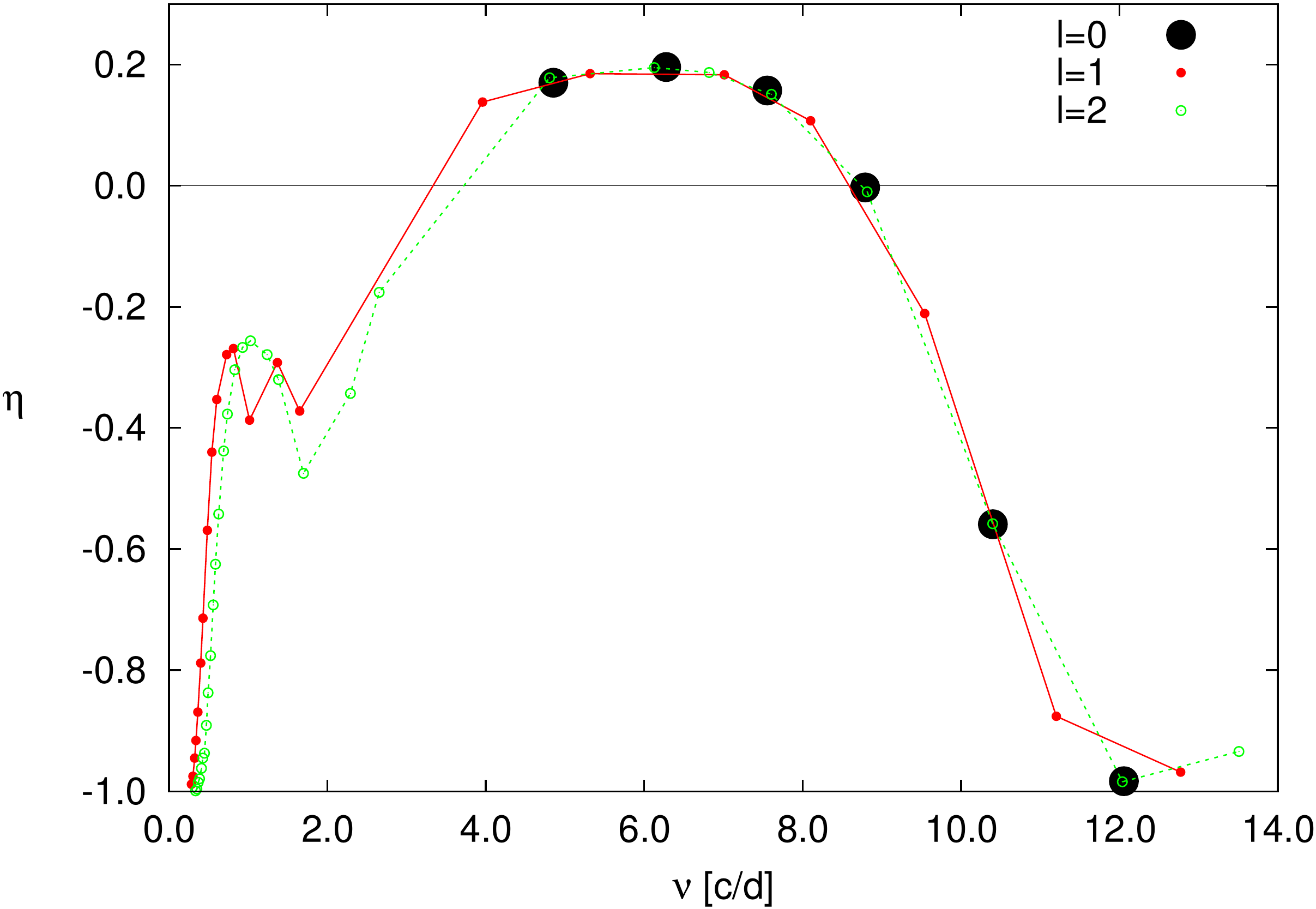}
   \caption{Instability parameter, $\eta$, as a function of frequency for $\ell=0,1,2$ for the post-MS (left panel) and MS model (right panel)
   with the mass of $16 M_\odot$  and $\log T_\mathrm{eff}=4.340$ and 4.445, respectively. The models were computed with the OP tables and AGSS09 mixture.}
\label{fig2}
\end{center}
\end{figure*}
\begin{figure*}
\begin{center}
 \includegraphics[clip,width=65mm,height=85mm,angle=-90]{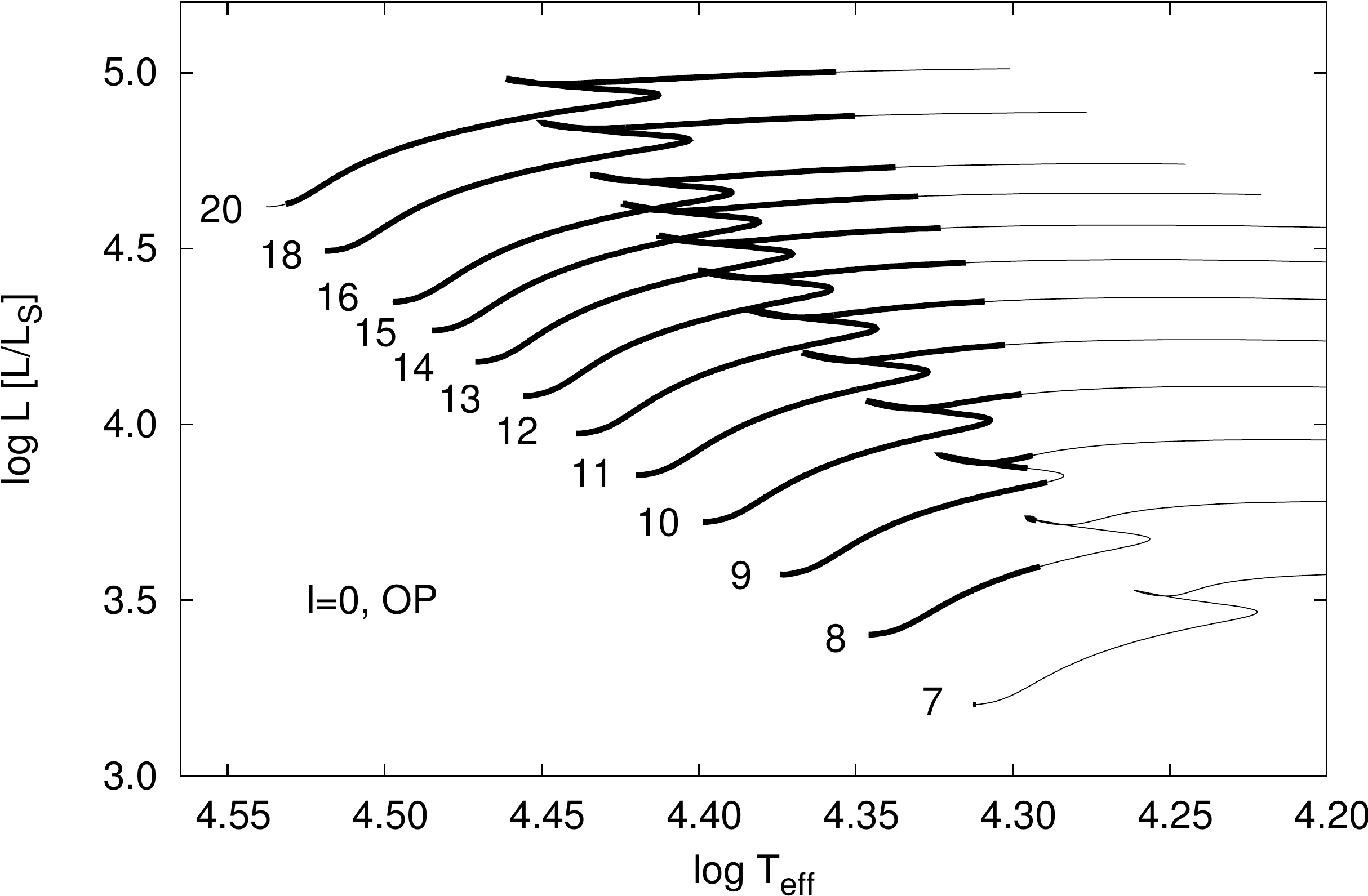}
 \includegraphics[clip,width=65mm,height=85mm,angle=-90]{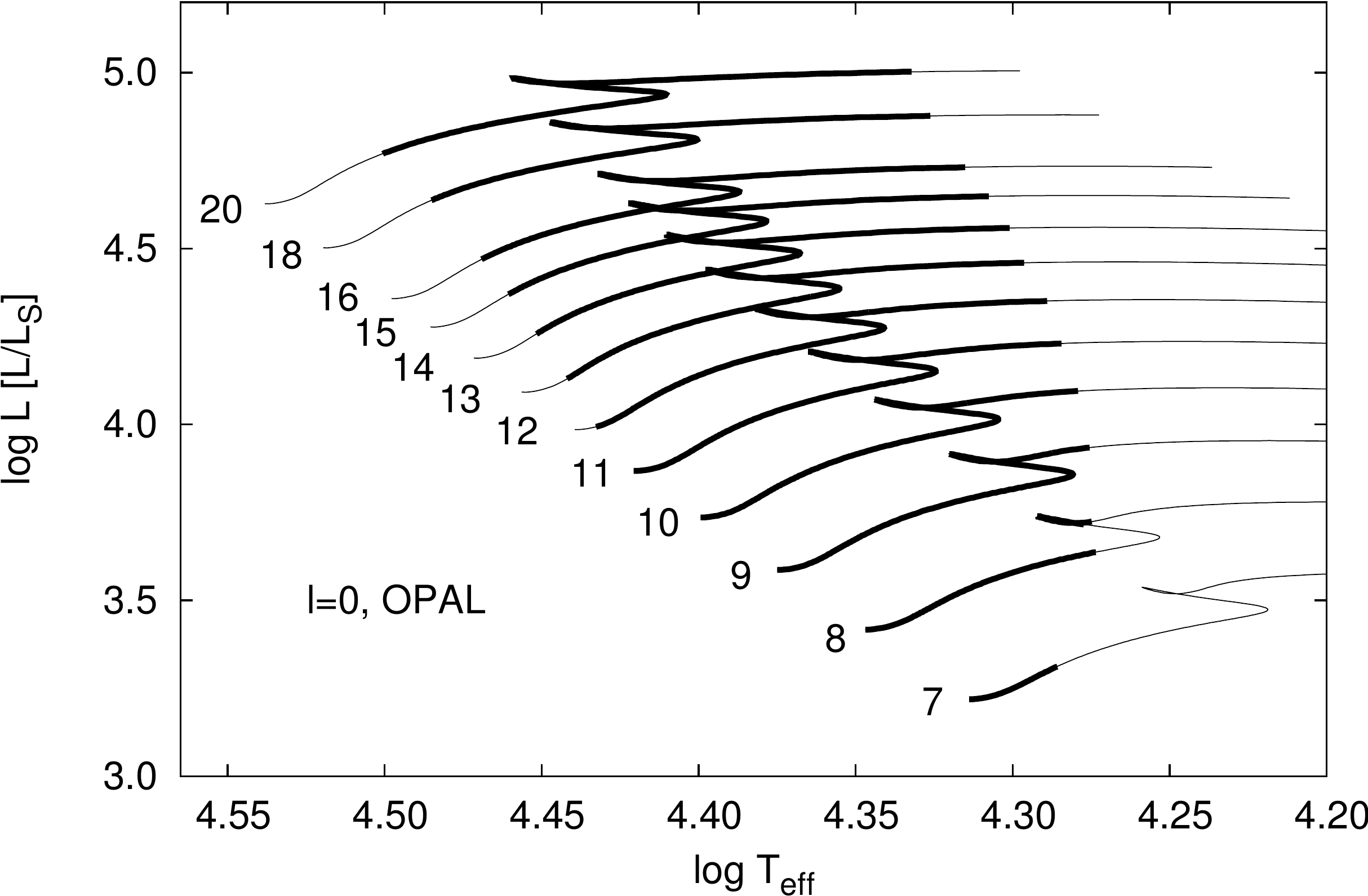}
 \includegraphics[clip,width=65mm,height=85mm,angle=-90]{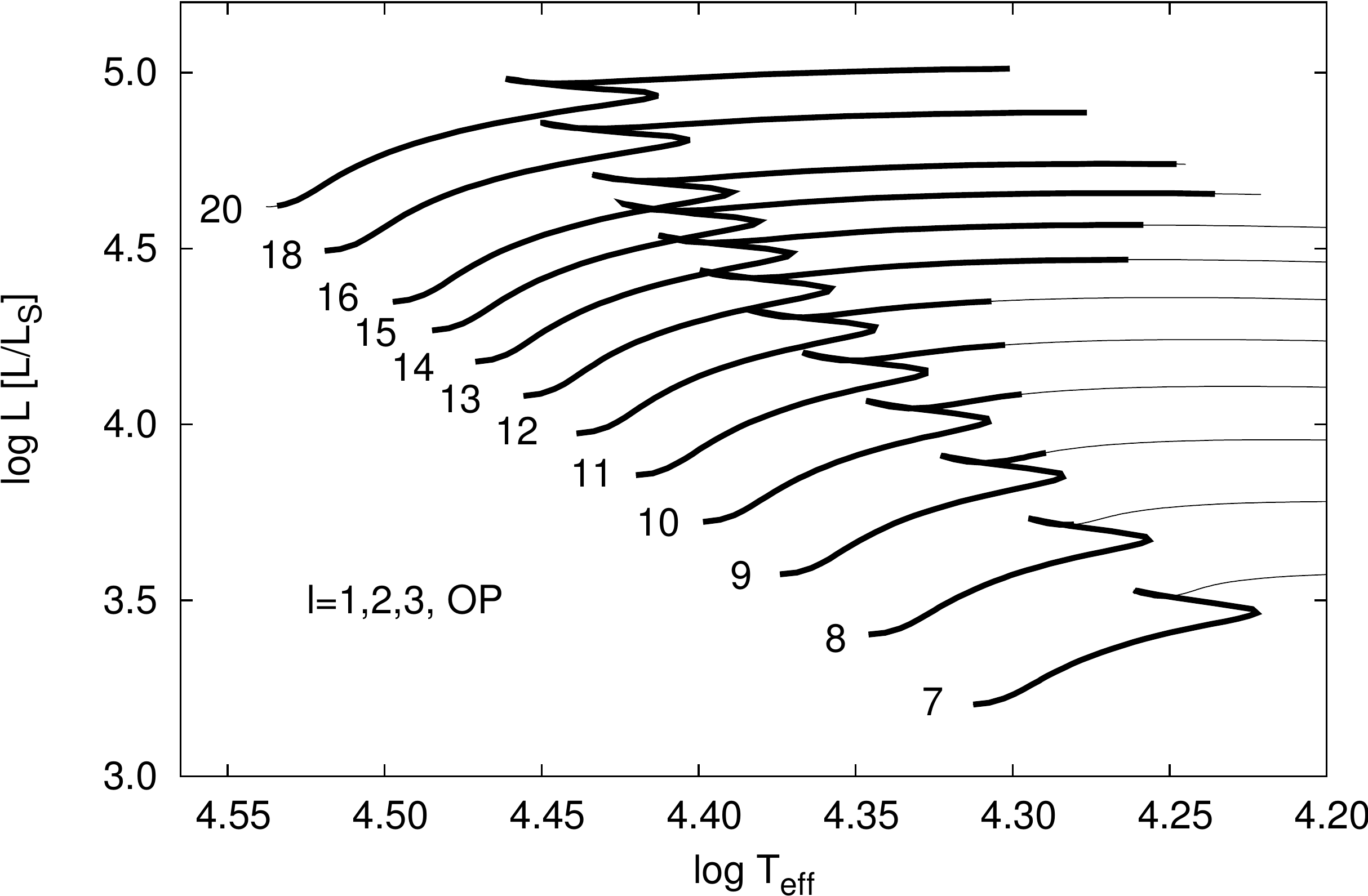}
 \includegraphics[clip,width=65mm,height=85mm,angle=-90]{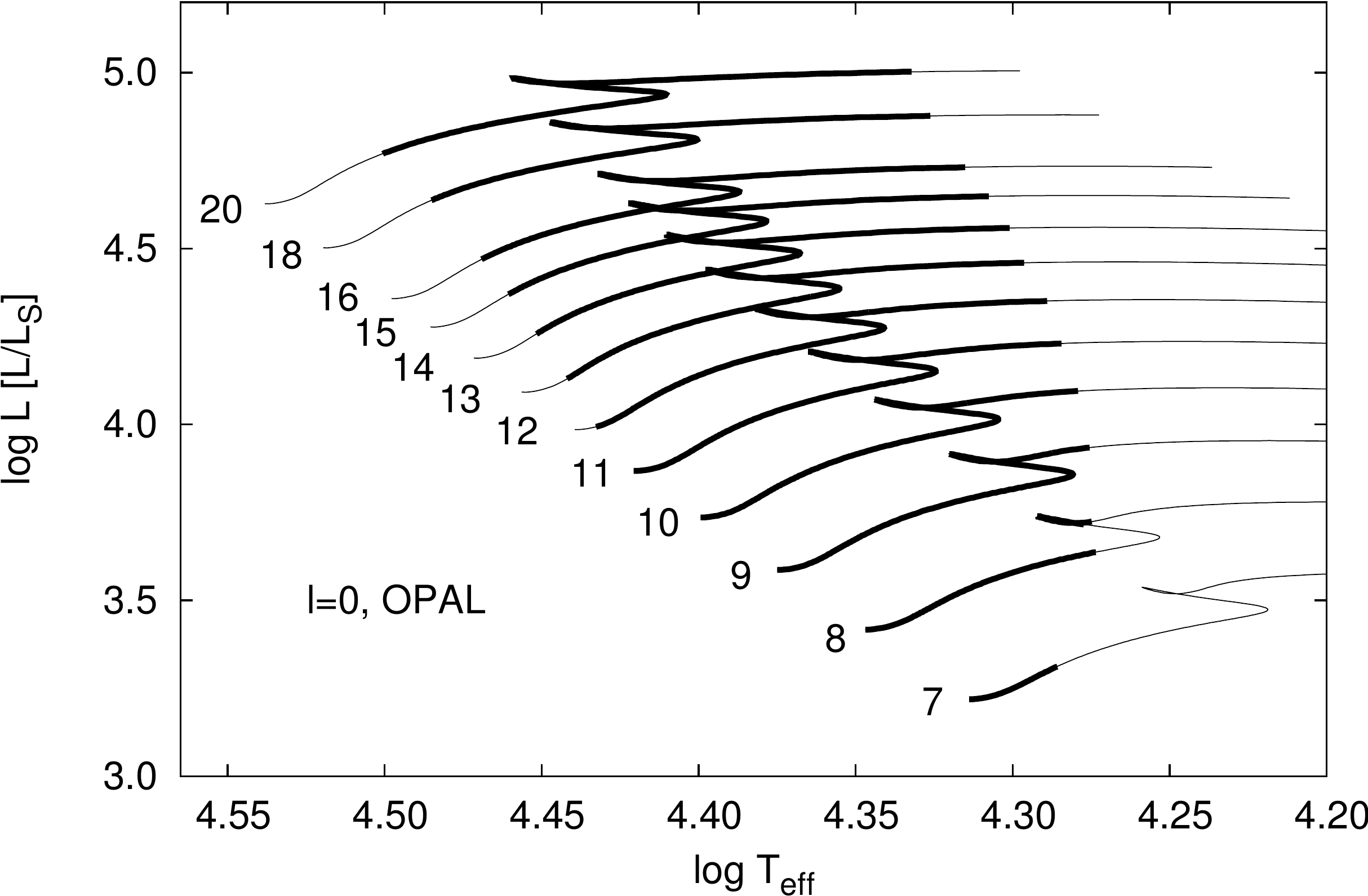}
   \caption{The HR diagram with instability domains, marked as thick lines, for the radial (upper panels) and nonradial (lower panels) modes of models with masses of $7-20 M_\odot$.
    The left panels show models with unstable modes computed with the OP opacities whereas the right panels - with the OPAL data. The AGSS09 chemical mixture,
    hydrogen and metal abundance of $X=0.7$ and $Z=0.02$, respectively, were assumed.}
\label{fig3}
\end{center}
\end{figure*}

In the next step, we have checked instability of supergiant models against radial and nonradial pulsation.
To this end, we used the non-adiabatic code of Dziembowski (1971, 1977).
As one can see in Fig.\,1, two widely separated regions of unstable modes are present. The one occurs at low frequencies (0.2-0.3 c/d) and corresponds to gravity modes
with very high radial orders, $n\approx 250-350$. The second one at frequencies 1.8-2 c/d is related with lower order g-modes ($n\approx 30-50$) which behave like p-modes
only in very outer layers and with a radial fundamental mode.
These results are presented also in the left panel of Fig.\,3 where we show the instability parameter, $\eta$, as a function of the mode frequency
for modes with $\ell=0,1$ and 2. For a comparison, in the right panel of Fig.\,3, we put a typical run of $\eta(\nu)$ for a main sequence model
with the same mass and $\log T_{\rm eff}=4.445$.
Let us recall that $\eta$  measures the net energy gained by a mode during one pulsational cycle and is defined as
$$\eta=\frac{W}{\int\limits_0^R\left|\frac{\mathrm{d}W}{\mathrm{d}r}\right|\mathrm{d}r}, \eqno(3)$$
where $W$ is the global work integral. If $\eta > 0$, driving overcomes damping and a mode is unstable.
As we can see the general dependencies of $\eta(\nu)$ for the post-main sequence (post-MS) model and main sequence (MS) model are similar, i.e.,
there are two global maxima of $\eta$. The first one is related to the high-order g-modes and occurs at $\nu\approx 0.9$ and 0.25 c/d
for the MS and post-MS model, respectively. The second global maximum corresponds to low order p/g-modes and occurs at $\nu\approx 6.2$ and 1.8-2.0 c/d
for the MS and post-MS model, respectively. The difference is that for the supergiant model, within the low frequency global maximum
there is an additional pattern of many local maxima and minima. as we have already mentioned, in the post-MS models
all nonradial modes associated with the higher frequency maximum have mostly gravity character;
they behave like p-modes only in the very outer envelope.

This behaviour of $\eta$ in the post-MS model can be explained by the effect of mode trapping. If there is a node of the radial displacement eigenfunction
in the vicinity of the $\mu$-gradient zone, a cavity is divided into two parts and a mode is trapped in the radiative helium core and in the envelope.
In the envelope a mode behaves like a $p$-mode and energy trapped there is sufficient for the opacity mechanism to be effective. In such a case, we have
the positive value of $\eta$.
The other situation is when there is no node in the vicinity of the $\mu$-gradient zone. In this case a mode enters the radiative core
with much higher amplitude and is very effectively damped. Then, a mode is stable.
Therefore, two very close frequency modes can have opposite energetic properties; the one is unstable whereas the other is stable.

We performed the instability analysis for models with masses from 7 $M_\odot$ up to 20 $M_\odot$, the same as considered by Saio et al. (2006).
Pulsational instability was found in post-MS models with masses greater than about 9 $M_\odot$ up to 20 $M_\odot$.
We did not consider models with masses higher than 20 $M_\odot$ because it would require taking into account the effects of mass loss.
As we did not consider models in the phase of the central helium burning, we evolved models until the temperature
in the center of a star reaches $10^8$ K, i.e., before the onset of the $3\alpha$ reactions.

In Fig.\,4, we show the instability domains on the HR diagrams marked as thick lines.
Models with unstable radial and nonradial modes are depicted separately in the upper and lower panels, respectively.
The left panels correspond to computations performed with the OP opacity tables and the right panels to those obtained with the OPAL data.
The main difference is that with the OP data unstable modes appear from the very beginning on main sequence, i.e., from ZAMS for all masses but $M=20 M_\odot$.
With the OPAL tables we got instability from ZAMS for $M\le 11 M_\odot$ and for higher masses it is shifted to the later phases on main sequence.
As for pulsational instability of post-MS models, the difference between computations with the OP and OPAL data is subtle; with the OPAL tables
the instability domain extends to slightly lower effective temperatures.

Our results differ somewhat from those obtained by Saio et al. (2006) who adopted the OPAL data. Unlike these authors we did not have gaps
in instability for the post-MS models with $M=11 -13 M_\odot$. Moreover, our instabilities do not extend to such low effective temperature
as about 14 000 K. This may be due to different chemical mixture adopted and/or differences in evolutionary and pulsational codes.

\begin{figure}
\begin{center}
 \includegraphics[clip,width=69mm,height=87mm, angle=-90]{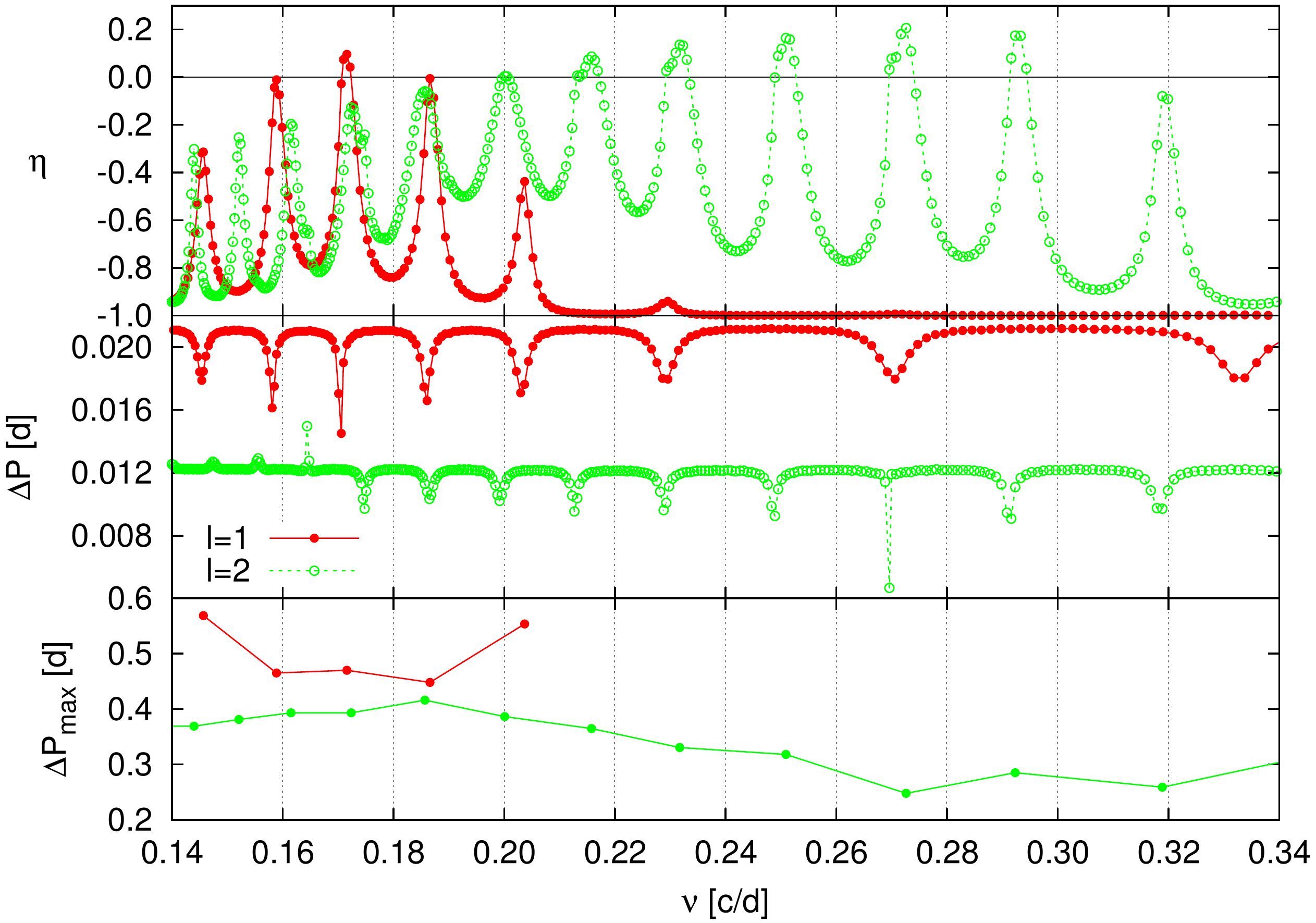}
   \caption{Correlation between the instability parameter, $\eta$ (top panel), period spacing, $\Delta P$ (middle panel) and $\Delta P_{\max}$ (bottom panel)
in the frequency range of high-order $g$-modes with $\ell=1$ and 2 in the post-MS OP model with M=16 $M_\odot$, $\log T_{\rm eff}=4.340$ and $\log L/L_\odot =4.730$.}
  \label{fig4}
\end{center}
\end{figure}

\begin{figure*}
\begin{center}
 \includegraphics[clip,width=70mm,height=105mm]{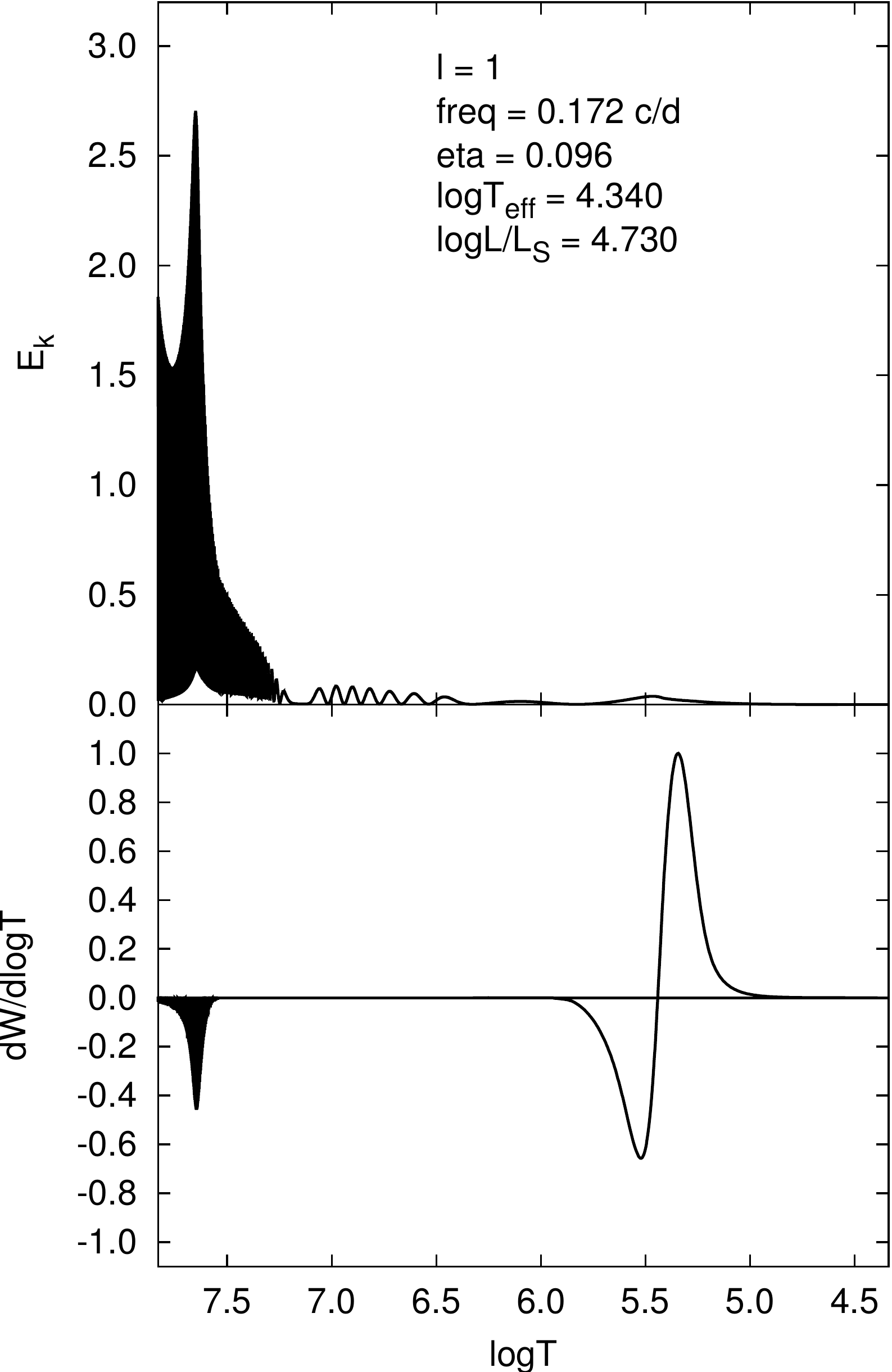}
 \includegraphics[clip,width=70mm,height=105mm]{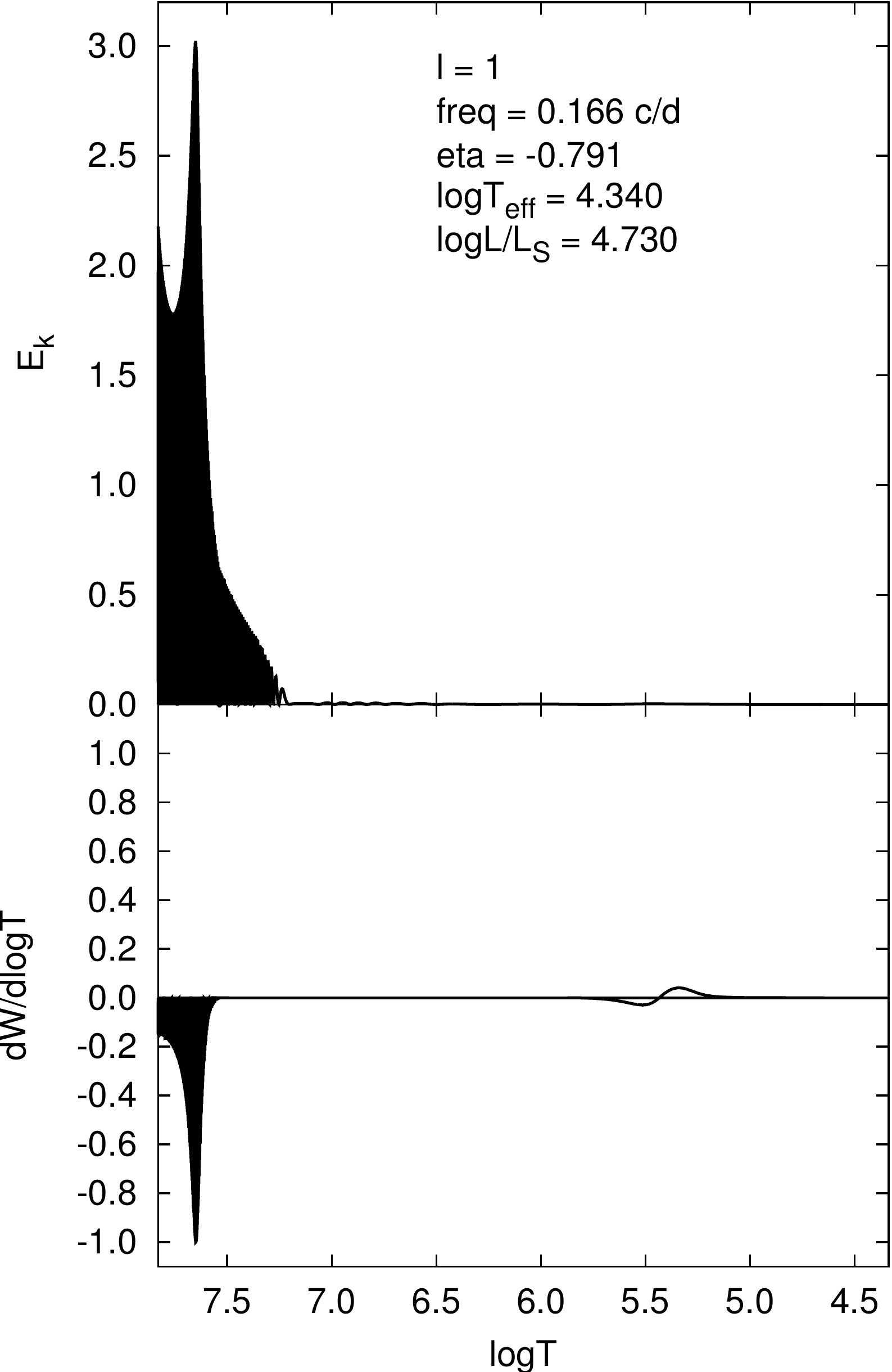}
   \caption{The kinetic energy density (top panels) and differential work integral (bottom panels) of unstable (left panels) and stable (right panels)
   $\ell=1$ modes with close frequencies. The values of $\nu$ and $\eta$ are listed in panels. The same OP model as in Fig.\,1 was used.}
\label{fig5}
\end{center}
\end{figure*}

\section{Properties of the SPBsg modes}
The asymptotic relations for high-order g-modes predict that at a given mode degree, $\ell$, periods of consecutive radial orders, $n$,
should be equidistant (Tassoul 1980). It is clearly visible in the middle panel of Fig.\,5 where the value of $\Delta P$, defined as
$$\Delta P = P_{n,\ell} - P_{n-1,\ell},\eqno(3)$$
is plotted as a function of the frequency, $\nu$, for $\ell=1$ and 2.
We can see that the period spacings are generally constant and amounts to $\Delta P\approx 0.021$ [d] for $\ell=1$ and 0.012 [d] for $\ell=2$,
but there are local minima occurring when the instability parameter reaches maxima (upper panel of Fig\,5).
This behaviour of $\Delta P$ and its correlation with $\eta$ is a consequence of mode trapping.

It is also interesting to note that the period spacing between consecutive maxima of $\eta(\nu)$ is not constant and amounts to $\Delta P_{\rm max}=0.45-0.57$ [d]
for $\ell=1$ and $\Delta P_{\rm max}=0.25-0.42$  for $\ell=2$ (the lower panel of Fig.\,5). These maxima correspond to eigenmodes of the outer g-cavity.
Therefore, in the case of oscillation spectra of the SPBsg pulsators one should look rather for irregularities corresponding to $\Delta P_{\rm max}$.

The phenomenon of mode trapping can be better explained by the distribution of the kinetic energy density, $E_k$, of modes and its radial and horizontal components.
We adopted the usual normalization $\int_0^1 E_k dx=1$, where $x=r/R$.
In Fig.\,6, we plotted the kinetic energy density (upper panels) and differential work integral (lower panels) for two very close frequency modes with $\ell=1$,
one of which is unstable (left panels) and the other stable (right panels).
The unstable mode has a frequency of 0.172 c/d and corresponds to the local maximum of $\eta$, whereas the stable mode, with a frequency of 0.166 c/d,
corresponds to the local minimum of $\eta$ (cf. Fig.\,5). We considered the same OP model as presented in Fig.\,1.
As we can see, the kinetic energy density is higher for the stable mode than for unstable one. For both modes the kinetic energy density
is almost entirely confined to the radiative core where damping is extremely effective, what can be estimated from the values of $dW/d\log T$
shown in the lower panels of Fig.\,6. The difference is that the unstable mode has some addition of $E_k$ in the outer layers.
It is more visible when one decomposes the kinetic energy density into the radial and horizontal components.

Fig.\,7 shows the radial (upper panels) and horizontal (bottom panels) components
of $E_k$ for the same unstable (left panels) and stable (right panels) high-order g-modes as considered in Fig.\,6.
The horizontal component, which is related to g-mode behaviour, is much higher than the radial component for all high-order g-modes in SPBsg stars
but the small radial component is crucial for instability. The radial component is related to the p-mode behaviour and it is contained mainly
in the envelope of the star. Modes perfectly trapped in both cavities have smaller total kinetic energy, hence the damping in the core
is smaller and the higher radial component of energy in the envelope makes the opacity driving mechanism effective. On the contrary, partially trapped
modes have very high energy in the core and not sufficient energy in the envelope in order for the $\kappa$-mechanism to be effective.

\begin{figure*}
\begin{center}
 \includegraphics[clip,width=115mm,height=130mm]{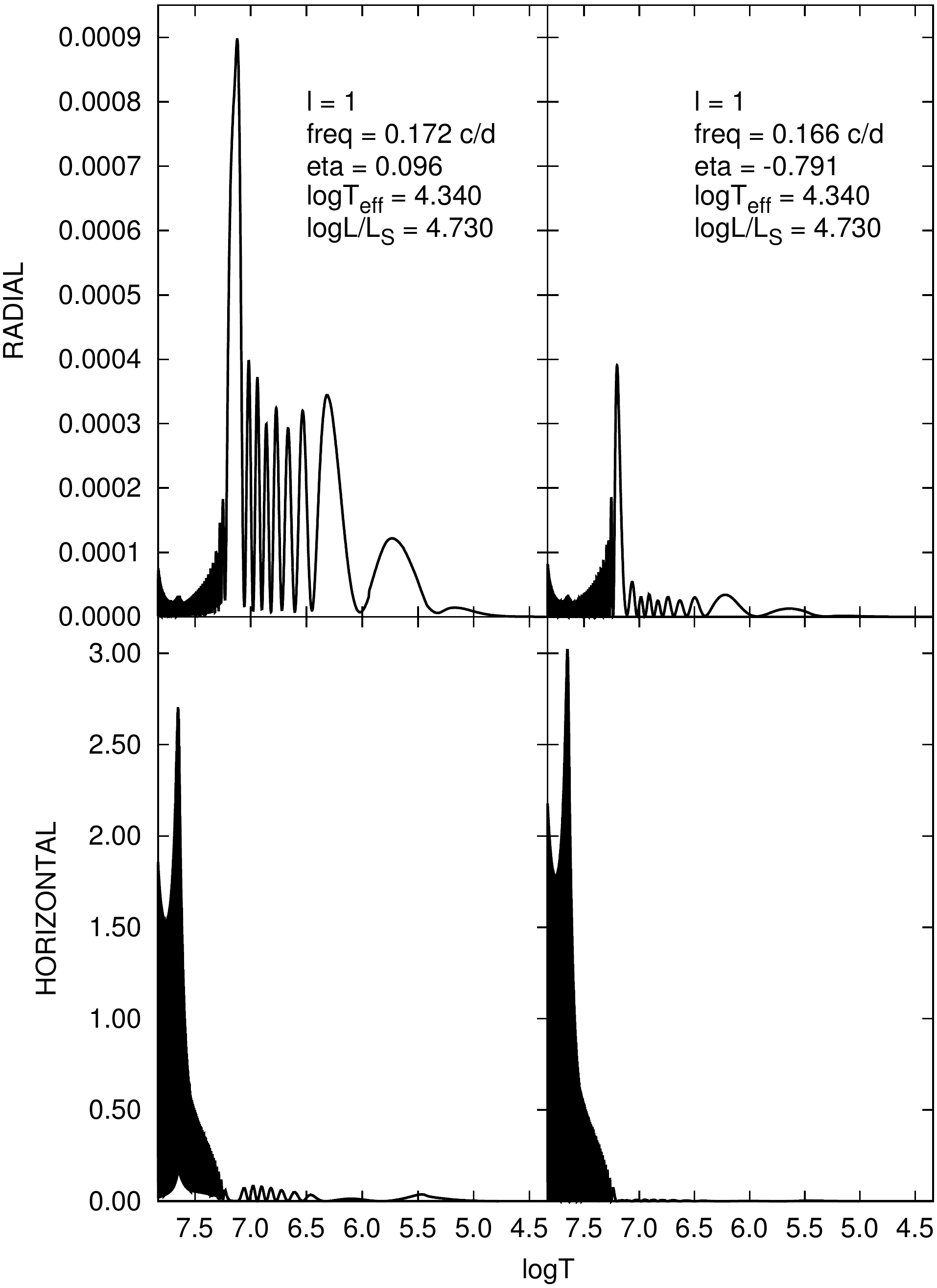}
   \caption{The radial (upper panels) and horizontal (bottom panels) components of the kinetic energy density for modes considered in Fig.\,5.}
  \label{fig7}
\end{center}
\end{figure*}

It is important to add, that partial reflection of some oscillation modes always occurs in those specific stellar layers where the structure is characterized
by sharp spatial changes. For gravity modes, the behavior of the Brunt-V\"ais\"al\"a frequency is critically important. In our case, the reflection of some g-modes occurs at the top of the hydrogen gradient zone where the Brunt-V\"ais\"al\"a frequency has a deep minimum and where actual temperature gradient is changed from the adiabatic value to the radiative one ($\log T \approx 7.14$ and  $r/R\approx 0.144$). Modes with small amplitudes (nodes) at this layer are reflected here.
In other words, a mode trapping occurs in two regions, i.e., below and above of this layer. Due to the very high value of $N$ in deep interior,
the oscillation spectrum is very dense, but due to quite small values of $N$ in outer part only few modes from such a dense spectrum (most of waves are confined in deep interior) have nodes close to the layer under consideration and are definitively trapped in two regions. The trapping in the outer part is critical for the instability of such modes.

This is shown in Fig. 8 which demonstrates the reflection (or trapping) of some modes at the layer where the Brunt-V\"ais\"al\"a frequency has deep minimum.
Again, two dipole gravity modes with close frequencies are considered. The trapped mode g$_{282}$ with the frequency 0.1711 c/d is unstable
and the untrapped mode g$_{270}$ with the frequency 0.1794 c/d is stable. From the top panel we can see that 17 \% of the kinetic energy of the trapped mode
is confined in the outer part (above minimum of $N$), whereas for the untrapped mode almost the whole kinetic energy (more than 99 \%) is confined in the deep interior
where this mode is strongly damped (see Fig.\,6 where stellar regions of damping and excitation are shown).
Note that with standard normalization used in linear computations (the amplitude of the radial displacement on the surface is assumed to be 1) the total kinetic energy of the trapped mode g$_{282}$ is 22 times less that those of untrapped mode g$_{270}$, therefore the g$_{282}$ mode can be excited much easier.
Lower panels illustrate the reflection (or mode trapping) phenomenon. The trapped mode is evanescent at the reflecting layer (which is illustrated in the small panel) and has high oscillation amplitudes above this layer whereas the untrapped mode is almost totally confined in the deep interior and has not evanescent behavior at the minimum of the Brunt-V\"ais\"al\"a frequency.

\begin{figure*}
\begin{center}
 \includegraphics[clip,width=115mm,height=130mm]{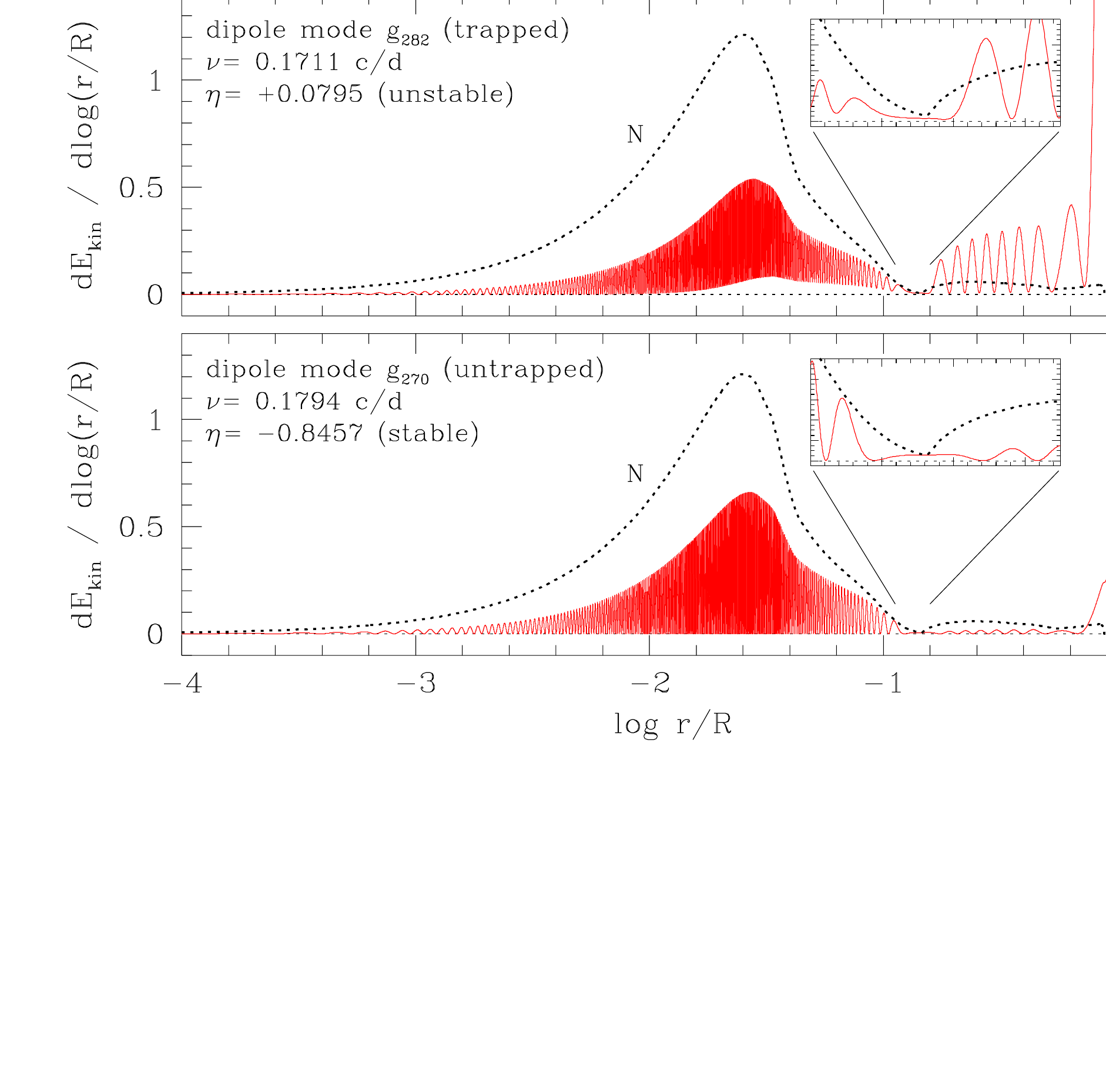}
   \caption{The normalized kinetic energy (top panel) and the kinetic energy density (lower panels) of two dipole modes with close frequencies.
   The trapped mode g$_{282}$ with the frequency 0.1711 c/d is unstable, and the untrapped mode g$_{270}$ with the frequency 0.1794 c/d is stable.
   The partial reflection (trapping) of g$_{282}$ mode occurs at the top of the hydrogen gradient zone (at $\log r/R = -0.840$, $r/R=0.144$, $\log T = 7.142$),
   where the Brunt-V\"ais\"al\"a frequency has a deep minimum (dotted line, in arbitrary units). The position of this minimum is marked by the vertical line.
   For the trapped mode, 17 \% of the kinetic energy is confined in the outer part above the reflection point.}
  \label{fig8}
\end{center}
\end{figure*}

\section{Prospect for mode identification}

As we could see, a number of modes can be excited in the B-type supergiant models.
Due to the specific behaviour of the instability parameter, $\eta$, (cf. Fig.\,5)
one should rather not expect a constant period spacing in oscillation spectra of this new emerged type of pulsators.
Therefore, let us now consider a possibility of identification of the SPBsg modes from multicolour photometry.

\begin{figure*}
\begin{center}
 \includegraphics[clip,width=67mm,height=85mm,angle=-90]{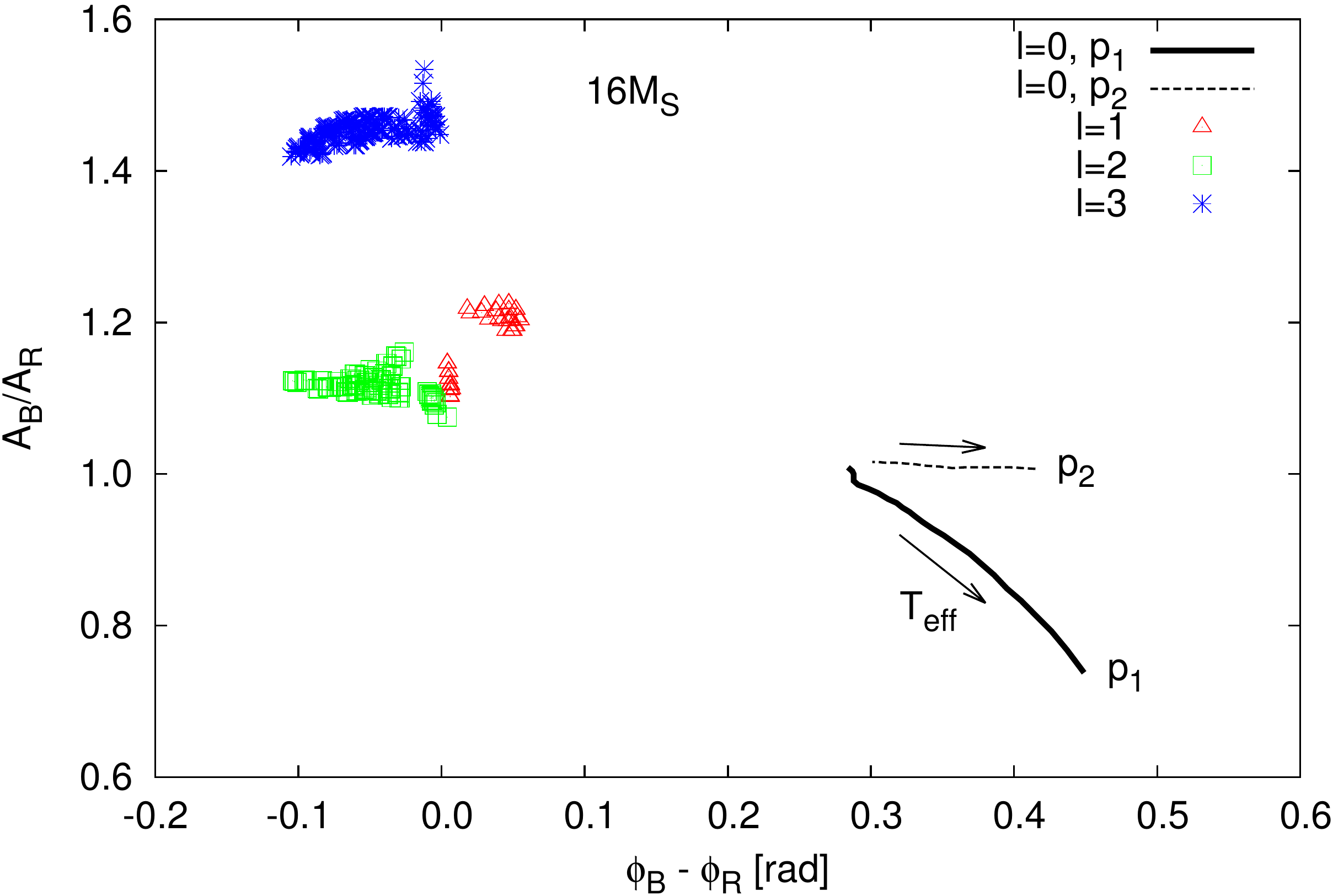}
 \includegraphics[clip,width=67mm,height=85mm,angle=-90]{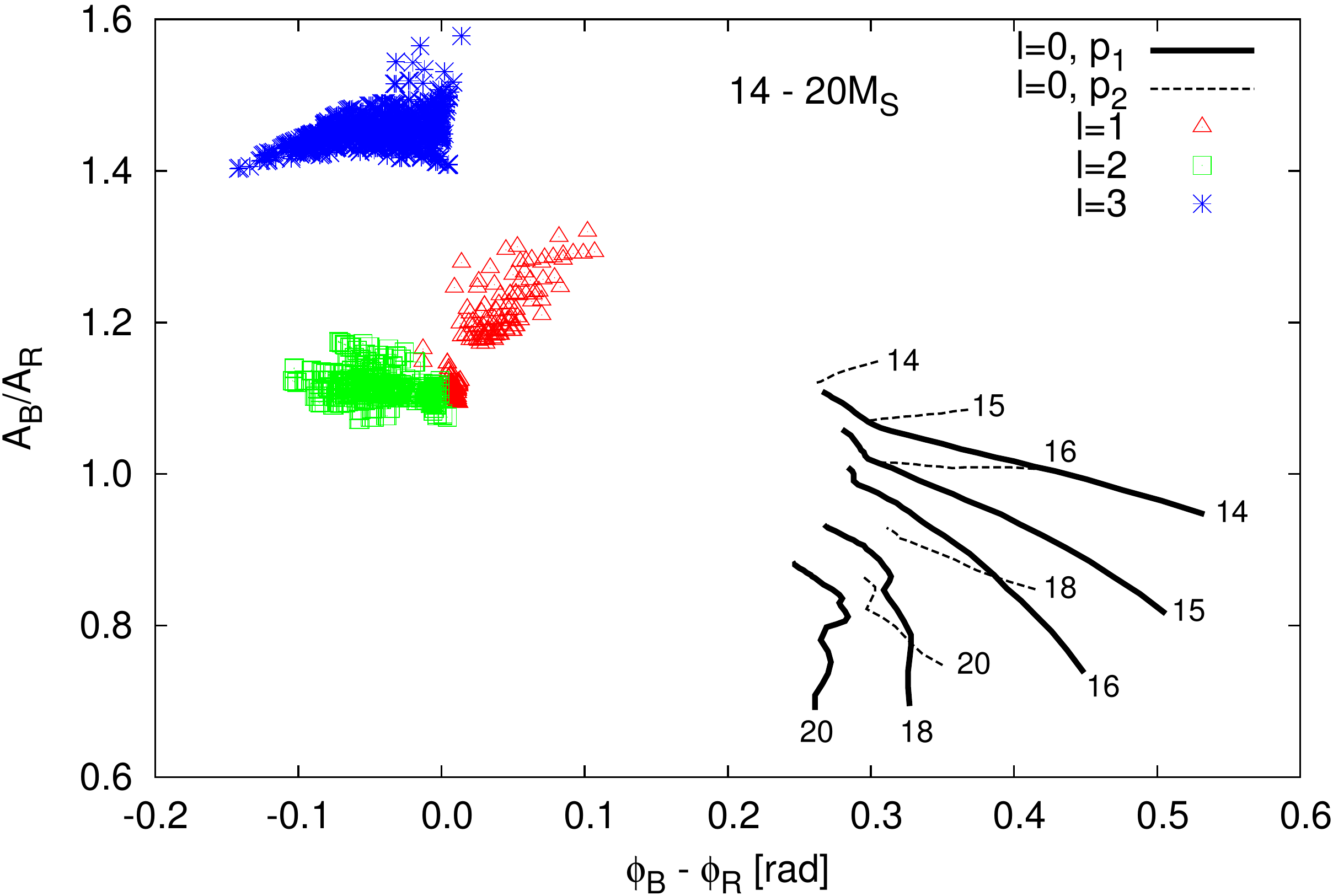}
 \caption{The photometric diagnostic diagrams in the Johnson BR passbands with positions of unstable modes with $\ell \le 3$
  found in the post main sequence models. The left panel includes unstable modes of the post-MS models with a mass of 16 $M_{\odot}$
  and the right panel - unstable modes of the post-MS models with masses of 14, 15, 16, 18 and 20 $M_\odot$.
  The OP opacities, AGSS09 mixture and NLTE model atmospheres were adopted.}
\label{fig9}
\end{center}
\end{figure*}

If all effects of rotation on pulsation can be ignored, the complex amplitude of the light variations
in the passband $\lambda$ for a star pulsating in a mode with the frequency $\omega$, the degree $\ell$
and azimuthal order $m$ is expressed as (e.g., Daszy\'nska-Daszkiewicz et al. 2002):
$${\cal A}_{\lambda}(i) = -1.086 \varepsilon Y_{\ell}^m(i,0) b_{\ell}^{\lambda}
(D_{1,\ell}^{\lambda}f+D_{2,\ell}+D_{3,\ell}^{\lambda})\eqno(1)$$
where
$$D_{1,\ell}^{\lambda} = \frac14  \frac{\partial \log ( {\cal
F}_\lambda |b_{\ell}^{\lambda}| ) } {\partial\log T_{\rm{eff}}} ,
\eqno(2a)$$
$$D_{2,\ell} = (2+\ell )(1-\ell ), \eqno(2b)$$
$$D_{3,\ell}^{\lambda}= -\left( 2+ \frac{\omega^2 R^3}{GM} \right)
 \frac{\partial \log ( {\cal F}_\lambda |b_{\ell}^{\lambda}|
) }{\partial\log g_{\rm eff}}, \eqno(2c)$$
Here, $\varepsilon$ is the intrinsic mode amplitude, $i$ is the inclination angle and  $G,M,R$ have their usual meanings.
The term $D_{1,\ell}^\lambda$ describes the temperature effects, $D_{2,\ell}$ - geometrical effects, and $D_{3,\ell}^\lambda$ - the influence
of the pressure changes. The terms $D_{1,\ell}^\lambda$ and $D_{3,\ell}^\lambda$ include the perturbation of the limb-darkening.
The disc-averaging factor, $b_{\ell}^{\lambda}$, is the integral of the limb-darkening weighted by the Legendre polynomial.
Derivatives of the monochromatic flux, ${\cal F}_\lambda(T_{\rm eff},\log g)$, are calculated usually from static, plane-parallel
atmosphere models. In general, these derivatives depend also on the metallicity parameter [m/H] and microturbulent velocity $\xi_t$. Here, we rely on
the NLTE atmosphere models of Lanz \& Hubeny (2007). We use the fluxes in photometric passbands and nonlinear limb-darkening coefficients
computed by Daszy\'nska-Daszkiewicz \& Szewczuk (2011).

The $f$ parameter describes the ratio of the bolometric flux perturbation to the radial displacement at the level of the photosphere:
$$\frac{ \delta {\cal F}_{\rm bol} } { {\cal F}_{\rm bol} }=
{\rm Re}\{ \varepsilon f Y_\ell^m(\theta,\varphi) {\rm e}^{-{\rm i}
\omega t} \}.\eqno(3)$$
The values of $f$ is complex and it is obtained from linear computations of stellar nonadiabatic pulsations.

The positions of unstable modes with the mode degree, $\ell$, up to 3, on the diagnostic diagram employing the $B,R$ Johnson filters
are presented in Fig.\,9. We chose these filters because they are close to those used on the BRITE-Constellation
(BRIght-star Target Explorer), the first forthcoming Austrian-Canadian-Polish space mission which will perform two-colour photometry.
Here, we show results obtained with the OP data.
In the left panel of Fig.\,9,  we put unstable modes for the post-main sequence models with a mass of $M=16M_\odot$ and effective temperatures
from $\log T_{\rm eff}=4.384$ to 4.235. As we can see, separation of modes with different values of $\ell$ is good. In the case of radial modes,
the fundamental and first overtone modes follow different paths.
This picture is not spoiled even if models in a wide range of masses are considered. In the right panel of Fig.\,9, we put modes excited
in models with masses $M=14,15,16, 18$ and 20 $M_\odot$.
The diagnostic photometric diagrams for other pairs of passbands look quite similar but a combination of the $BR$ filters gives one of the best separation of modes.

Using the OPAL data we get similar locations of modes with different mode degrees. The only difference is that there are less unstable high-order
g-modes modes because of the smaller maximum of $\eta(\nu)$ in the lower frequency for the OPAL pulsational models.

\section{Conclusions}
The goal of this paper was to analyse the pulsational instability of the B-type supergiant stellar models and to examine properties of their oscillation modes.
We have showed that pulsational modes in the post-MS massive star models can be reflected at the chemical composition gradient zone developed above the radiative helium core
and then the excitation in the $Z-$bump is effective. Both radial and nonradial modes are present.
Our results are in contradiction with the results obtained by Saio et al. (2006) and Godart et al. (2009) who found
that a presence of an intermediate convective zone (ICZ) in the hydrogen-burning shell is a necessary condition to get unstable modes in these hot evolved stars.

As in the case of main sequence pulsators, the instability parameter, $\eta$, of post-MS models has maxima both at low and higher frequencies.
In the low frequency range, the values of $\eta$ reach many local maxima and minima which correlate with the period spacing.
The two modes with very close frequencies can differ in their kinetic energy and stability properties. The kinetic energy and its distribution inside the star
is different for unstable and stable high-order g-modes. For instability, the radial component of $E_k$ is crucial.
These properties of the SPBsg modes are consequences of the nature of mode trapping in two cavities of the star, i.e.,
in the radiative helium core and in outer envelope.

We obtained that equidistant pattern in the oscillation spectra of the SPBsg pulsators is rather not expected.
This fact prompted us to explore a possibility of identification of the SPBsg modes from the multicolour photometry.
We showed that the SPBsg modes with different degree $\ell$ are located in well separated regions on the photometric diagnostic diagrams.
In particular, a very good separation was obtained for a pair of the $BR$ Johnson passbands which are quite similar
to those applied on the BRITE satellites. Therefore, a great hope is pinned on this forthcoming space project.

\section*{Acknowledgments}
We gratefully thank Wojtek Dziembowski for helpful comments and advices. The work was partially financial supported
by the Polish NCN grants 2011/01/B/ST9/05448 and 2011/01/M/ST9/05914.

\end{document}